\documentclass[preprint,superscriptaddress]{revtex4-1}

\usepackage{lineno,hyperref,amsmath,graphicx,subfigure}

\newcommand{\lef}{\left(}
\newcommand{\rig}{\right)}
\newcommand{\deriv}{\textnormal{d}}

\begin{document}

\title{Resistivity scaling in metallic thin films and nanowires due to grain boundary and surface roughness scattering}

\author{Kristof Moors}
\email[E-mail me at: ]{kristof@itf.fys.kuleuven.be}
\affiliation{KU Leuven, Institute for Theoretical Physics, Celestijnenlaan 200D, B-3001 Leuven, Belgium}
\affiliation{Imec, Kapeldreef 75, B-3001 Leuven, Belgium}
\author{Bart Sor\'ee}
\affiliation{Imec, Kapeldreef 75, B-3001 Leuven, Belgium}
\affiliation{University of Antwerp, Physics Department, Groenenborgerlaan 171, B-2020 Antwerpen, Belgium}
\affiliation{KU Leuven, Electrical Engineering (ESAT) Department, Kasteelpark Arenberg 10, B-3001 Leuven, Belgium}
\author{Wim Magnus}
\affiliation{Imec, Kapeldreef 75, B-3001 Leuven, Belgium}
\affiliation{University of Antwerp, Physics Department, Groenenborgerlaan 171, B-2020 Antwerpen, Belgium}

\date{\today}

\begin{abstract}
A modeling approach, based on an analytical solution of the
semiclassical multi-subband Boltzmann transport equation,
is presented to study resistivity scaling in metallic thin
films and nanowires due to grain boundary and surface
roughness scattering. While taking into account the detailed
statistical properties of grains, roughness and barrier
material as well as the metallic band structure and quantum
mechanical aspects of scattering and confinement, the model
does not rely on phenomenological fitting parameters.
\end{abstract}

\maketitle

\section{Introduction} \label{sec:introduction}
The resistivity of metallic thin films and nanowires increases
drastically when the film thickness or wire diameter is reduced
\cite{josell2009size}. An increased resistivity is undesirable
for typical applications of these structures, e.g. interconnects
in semiconductor devices, as it leads to increased heating,
power dissipation, signal propagation delays, et cetera. Hence,
in order to assess the performance of metallic thin films and
nanowires as conductors in nanoscaled applications, it is
important to study their resistivity and scaling behavior and
understand how a drastic increase of resistivity can be
prevented, if at all possible for metallic structures with
sub-10 nm dimensions.

Experimental data has indicated that the increase of resistivity
is mainly induced by an increase of electron scattering at the
grain boundaries and near the rough boundaries of the structure.
These scattering mechanisms lead to a resistivity contribution
that adds to the bulk resistivity dominated by the electron-phonon
interaction and scattering with lattice imperfections which is,
to a good approximation, independent of the thickness. The
resistivity data of metallic thin films and wires is in good
agreement with the semiclassical Mayadas-Shatzkes model, commonly
used for data comparison and predicting a resistivity scaling
almost inversely proportional to the film width or wire diameter
\cite{josell2009size, mayadas1970electrical}. While the
Mayadas-Shatzkes model provides satisfactory fits to the data,
it contains phenomenological fitting parameters: a specularity
parameter for boundary surface scattering and a reflection
coefficient for grain boundary scattering. These parameters do
not provide a clear connection between the microscopic scattering
events and the resulting, measured resistivity of the thin film
or nanowire. For example, there is no clear relation between
boundary roughness, the microscopic origin of diffusive scattering
at the boundary, and the phenomenological specularity parameter
in the Mayadas-Shatzkes model which intends to capture this
process. Moreover, the Mayadas-Shatzkes model neglects the
material band structure properties and quantum mechanical aspects
of scattering and confinement while a priori there is no reason
to expect that both aspects have negligible impact on the
resistivity scaling behavior.

We present an alternative approach to model resistivity scaling
in metallic thin films and nanowires, based on the multi-subband
Boltzmann transport equation, with averaged scattering rates
obtained from Fermi's golden rule for grain boundary and surface
roughness scattering \cite{moors2014resistivity, moors2015modeling}.
Our approach allows to perform a rigorous analysis of the
resistivity and its scaling behavior while taking into account
the aforementioned aspects that are neglected in some
conventional approaches.

In section~\ref{section:theory} we summarize briefly the theory
of the semiclassical multi-subband Boltzmann equation and the
scattering rates obtained with Fermi's golden rule for grain
boundary and surface roughness scattering. Next, we present
some simulation results in section~\ref{section:results}, which
are discussed in section~\ref{section:discussion}, followed
by a conclusion in section~\ref{section:conclusion}. We also
refer to some articles with similar developments for metallic
thin films and nanowires \cite{fishman1989surface,
fishman1991influence,feilhauer2011quantum}.

\section{Theory} \label{section:theory}
The electron (or hole) transport formalism based on the
semiclassical multi-subband Boltzmann transport equation can
be summarized by the following list of equations:
\begin{align} \label{eq:BTE1}
	\mathbf{J} &= \sum_{\mathbf{n}} \int \mkern-3mu
		\frac{\deriv^D k}{(2 \pi)^D} \,
		\frac{q \, \mathbf{\nabla}_\mathbf{k}
		E_\mathbf{n} (\mathbf{k})}{\hbar} \,
		\delta f_\mathbf{n} (\mathbf{k}), \\
	\label{eq:BTE2} \delta f_\mathbf{n} (\mathbf{k})
		&= \frac{q \, \mathbf{E} \cdot
			\nabla_\mathbf{k}
			E_\mathbf{n}(\mathbf{k})}{\hbar}
		\, \tau_\mathbf{n} (\mathbf{k}) \,
		\delta \lef E_\mathbf{n}
			(\mathbf{k}) - E_\mathrm{F} \rig, \\
	\label{eq:BTE3} \frac{1}{\tau_\mathbf{n} (\mathbf{k})}
		&= \sum_{\mathbf{n}',\mathbf{k}'}
		\lef 1 - \frac{\tau_{\mathbf{n}'} (\mathbf{k}')
			\, \partial_{k'}
			E_{\mathbf{n}'}(\mathbf{k}')}
			{\tau_\mathbf{n} (\mathbf{k}) \, \partial_k
			E_\mathbf{n} (\mathbf{k})} \rig
		P\lef \mid \mathbf{n} \; \mathbf{k} \rangle
			\rightarrow \mid \mathbf{n}' \;
			\mathbf{k}' \rangle \rig, \\ \label{eq:BTE4}
	P\lef \mid i \rangle \rightarrow \mid f \rangle \rig &=
		\frac{2\pi}{\hbar}\left| \langle i \mid V \mid
			f \rangle \right|^2
			\delta\lef E_i - E_f \rig,
\end{align}
where $\delta f_\mathbf{n}$ is the deviation of the distribution
function from Fermi-Dirac equilibrium ($\delta f_\mathbf{n}(k)
\equiv f_\mathbf{n}(\mathbf{k}) - f^\textrm{FD}_\mathbf{n}
(\mathbf{k})$) for the (sub)band labeled by $\mathbf{n}$,
$E_\mathbf{n}(\mathbf{k})$ and $\tau_\mathbf{n} (\mathbf{k})$
are respectively the energy and relaxation time for a state
with wavevector $\mathbf{k}$ (and $k$ the component along the
direction of the electric field) in (sub)band $\mathbf{n}$, $q$
is the electron charge, $\mathbf{E}$ the electric field,
$E_\mathrm{F}$ the Fermi energy, $V$ the scattering potential
and $\mathbf{J}$ the current density. The dimensionality of
$\mathbf{n}$ and $\mathbf{k}$ depends on the system under
consideration. The wavevectors $\mathbf{k}$ are one-dimensional
($D=1$) for nanowires and two-dimensional ($D=2$) for thin films,
while $\mathbf{n}$ is a two-dimensional subband index vector for
two-dimensional nanowire confinement and one-dimensional for
thin film confinement (including an extra band index in both
cases if required). The list of equations follows from the
solution of the linearized Boltzmann equation at zero
temperature \cite{jacoboni2010theory}. The linearization and
zero temperature assumption are justified in the case of small
electric fields, elastic scattering and low enough temperatures
($k_\textrm{B} T \ll E_\textrm{F}$, with $E_\textrm{F}$ measured
from the lowest conduction band) and these are very reasonable
assumptions for typical metallic nanowires and thin films at
room temperature with electrons predominantly subjected to
grain boundary and surface roughness scattering. All the states
$\mid \mathbf{n}^( {}' {}^) \; \mathbf{k}^( {}' {}^) \rangle$
that are considered in Eq.~\ref{eq:BTE1}-\ref{eq:BTE4} are
therefore Fermi level states with $E_\mathbf{n} (k) =
E_{\mathbf{n}'} (k') = E_\textrm{F}$.

The relaxation times in Eq.~\ref{eq:BTE2} are coupled
self-consistently through a system of linear equations and can
be obtained through a matrix (of finite size for a nanowire
while requiring numerical discretization of $\mathbf{k}$ for
thin films) inversion. Fermi's golden rule is invoked to
obtain the scattering rates between the different electron
states due to grain boundary and boundary surface roughness
scattering. These scattering rates are averaged over an
ensemble of grain boundaries and surface roughness profiles
to retrieve a general and analytical expression which can be
inserted into Eq.~\ref{eq:BTE3}, allowing for fast and
accurate simulations. Because electron-phonon and imperfection
(e.g. point defects or impurities) scattering in thin films
and nanowires do not deviate much from their bulk scattering
behavior while being isotropic and independent from grain
boundary and surface roughness scattering (Matthiessen's
rule), their resistivity contribution is very close to the
bulk value, $\rho^\textrm{bulk}$, and can be separated from
the scaling part due to grain boundaries and surface roughness,
$\rho^\textrm{scaling}$. This consideration leads to a total
resistivity $\rho^\textrm{bulk} + \rho^\textrm{scaling}$,
with $\rho^\textrm{bulk}$ the bulk resistivity extracted
from experiments and $\rho^\textrm{scaling}$ resulting from
the solution of Eqs.~\ref{eq:BTE1}-\ref{eq:BTE4}.

The input which is required to solve
Eqs.~\ref{eq:BTE1}-\ref{eq:BTE4} consists of a correct band
structure profile of the nanowire or thin film, to be used
in Eqs.~\ref{eq:BTE1}-\ref{eq:BTE3}, the wave functions of
the electron states close to the Fermi level and expressions
for the grain boundary and surface roughness potentials,
entering the matrix elements in Eq.~\ref{eq:BTE4}. The set
of equations has no remaining free fitting parameters and
the resistivity can be obtained without numerical integration.

For grain boundaries, we have borrowed the scattering
potential and its distribution from the Mayadas-Shatzkes
model \cite{mayadas1970electrical}:
\begin{align} \label{eq:GB1}
	V^\textrm{GB}(x,y,z) &= \sum_{\alpha=1}^N
		S^\textrm{\tiny GB} \,
		\delta\lef z - z_\alpha \rig, \\ \label{eq:GB2}
	g\lef z_1, \ldots, z_N \rig &= \frac{\exp \left[-
		\sum\limits_\alpha (z_{\alpha+1} - z_\alpha -
		D^\textrm{\tiny GB})^2/ 2
		(\sigma^\textrm{\tiny GB})^2 \right]}
		{L_z [ 2\pi
		(\sigma^\textrm{\tiny GB})^2 ]^{(N-1)/2}},
\end{align}
where the grain boundaries are represented by $N$ Dirac delta
barrier planes normal to the transport ($z$) direction at
positions $z_\alpha$, the barrier strength $S^\textrm{\tiny GB}$
being distributed along the wire with an average distance
$D^\textrm{\tiny GB}$ in between subsequent grain boundaries
and standard deviation $\sigma^\textrm{\tiny GB}$. The average
distance and standard deviation can be estimated from the
experimental grain distribution while the barrier strength
(having units of energy times length), representing the
height and width of the grain boundary potential barrier,
can be extracted from \textit{ab initio} simulations. It
typically depends on the orientation of the grains and their
boundaries, but gives values of the order of magnitude of
eV$\cdot$\r{A}. The normal orientation of the grain boundary
planes in the Mayadas-Shatzkes model can be extended to random
orientations but the deviations in resistivity from the results
of grain boundaries with normal orientation are quite small
\cite{moors2015electron}.

For surface roughness, we consider the following potential
and statistics, based on Ando's surface roughness scattering
model \cite{ando1982electronic}:
\begin{align} \label{eq:SR1}
	V^\textrm{SR} (\mathbf{r}) &= U\lef x -
		\Delta(\mathbf{R}), y, z \rig - U
		\lef \mathbf{r} \rig, \\ \label{eq:SR2}
	\left\langle \Delta (\mathbf{R}) \right\rangle &=
		0, \quad \left\langle \Delta (\mathbf{R})
		\Delta (\mathbf{R}') \right\rangle =
		\Delta^2 e^{-(\mathbf{R} - \mathbf{R}')^2
			/(\Lambda^2/2) },
\end{align}
where $\mathbf{r} \equiv (x,y,z)$ and we assume a roughness
function $\Delta(\mathbf{R})$ with $\mathbf{R} \equiv (y,z)$
that shifts the potential $U(\mathbf{r})$ along a confinement
($x$) direction as a function of the boundary position
$\mathbf{R}$ with zero average, standard deviation (or RMS)
$\Delta$ and correlation length $\Lambda$. The matrix element
is linear in $V$ but not linear in $\Delta$. One often expands
the matrix element linearly in the roughness function in
combination with considering an infinite potential well for
$U(\mathbf{r})$, leading to the so called Prange-Nee
approximation for surface roughness scattering
\cite{prange1968quantum}. This approximation neglects the
oscillatory behavior of the wave functions and can lead to
large errors on the scattering rates. We have recently
introduced an analytical expression for the matrix elements
going beyond the linear expansion restriction as well as the
infinite potential well limit, hence avoiding additional
approximations such as the commonly used Prange-Nee
approximation \cite{moors2015modeling}. In this way, the
potential barrier outside the wire or film can also be
adjusted to represent the surrounding barrier material
accurately, improving once again the accuracy of the
simulations. While the roughness RMS and correlation length
can both be measured experimentally, the correlation length
is often neglected as it requires high resolution surface
imaging. A finite and accurate value of the correlation
length could be very important for nanowires however, as it
can facilitate the search for new types of state protection
from backscattering that may improve the resistivity.

\begin{figure}[tbh]
	\begin{center}
	\includegraphics[width=0.5\linewidth]{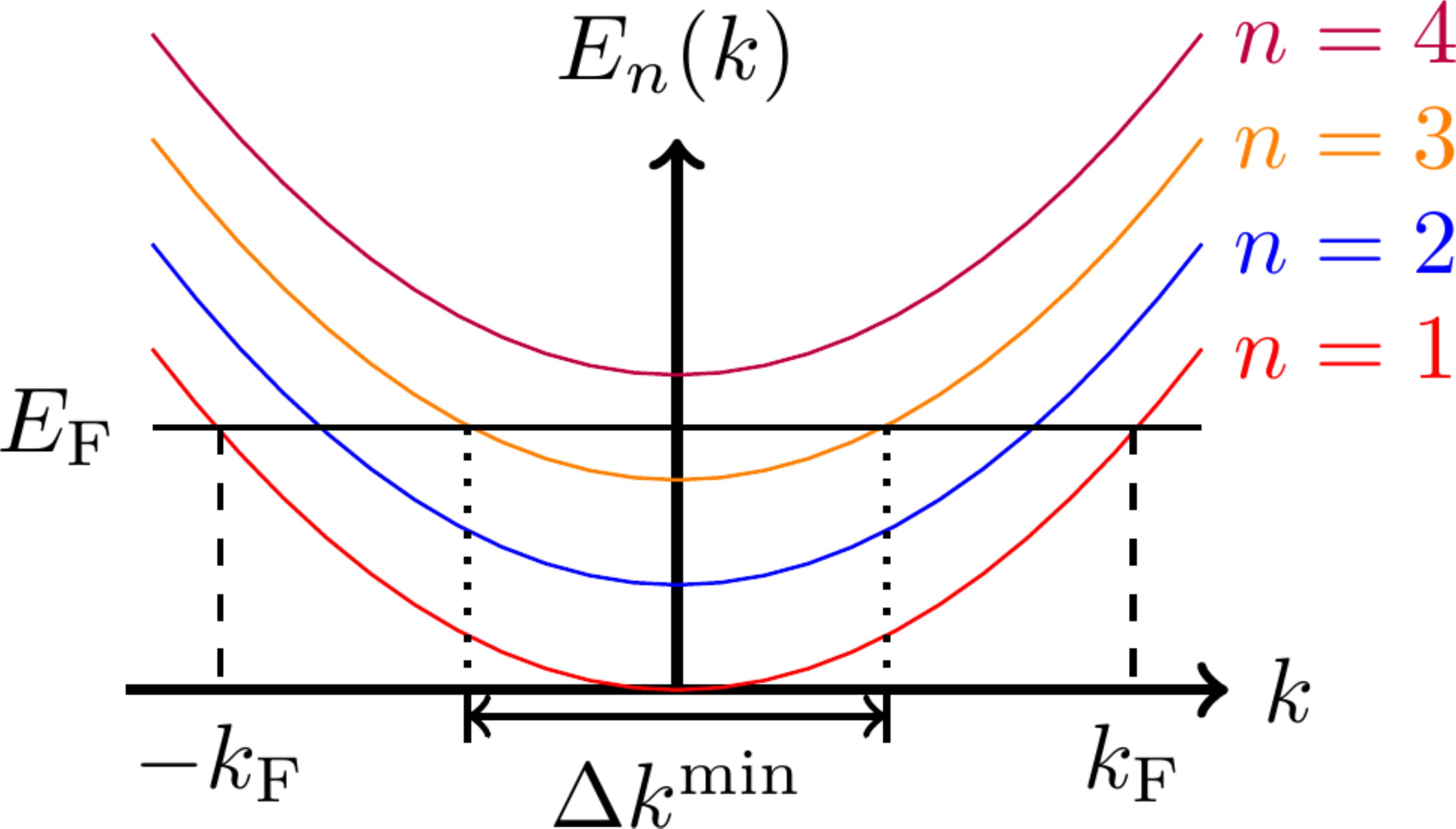}
	\end{center}
	\caption{A toy model band structure as a function of wave
	number $k$ along the transport direction corresponding to
	an effective mass description of the electrons with
	(sub)bands labeled by integer $n$. The Fermi (maximal)
	wave number $k_\textrm{F}$ is indicated as well as the
	minimal difference between positive and negative wave
	numbers $\Delta k^\textrm{min}$ at the Fermi energy level
	$E_\textrm{F}$.}
	\label{fig:Bands}
\end{figure}

\section{Results}
\label{section:results}

We will present the scattering properties, based on the scattering
potentials and statistics of Eq.~\ref{eq:GB1}-\ref{eq:SR2}, in the
first subsection and the corresponding resistivity results for
thin films and nanowires in the following subsection. For the sake
of simplicity, the results are limited to thin films and nanowires
represented by a finite potential well confining the electrons
that are described in the effective mass approximation (see
Fig.~\ref{fig:Bands}), although the present approach is generally
applicable. The effective mass $m_e^*$ and conduction electron
density $n_e$ are chosen to those of Cu: $m_e^*\approx m_e$,
$n_e \approx 8.469 \times 10^{28}$~m${}^{-3}$, $a_\textrm{Cu}
\approx 0.361$~nm.

\subsection{Scattering}
\begin{figure}[tbh]
	\centering
	\subfigure[]{\includegraphics[width=0.49\linewidth]{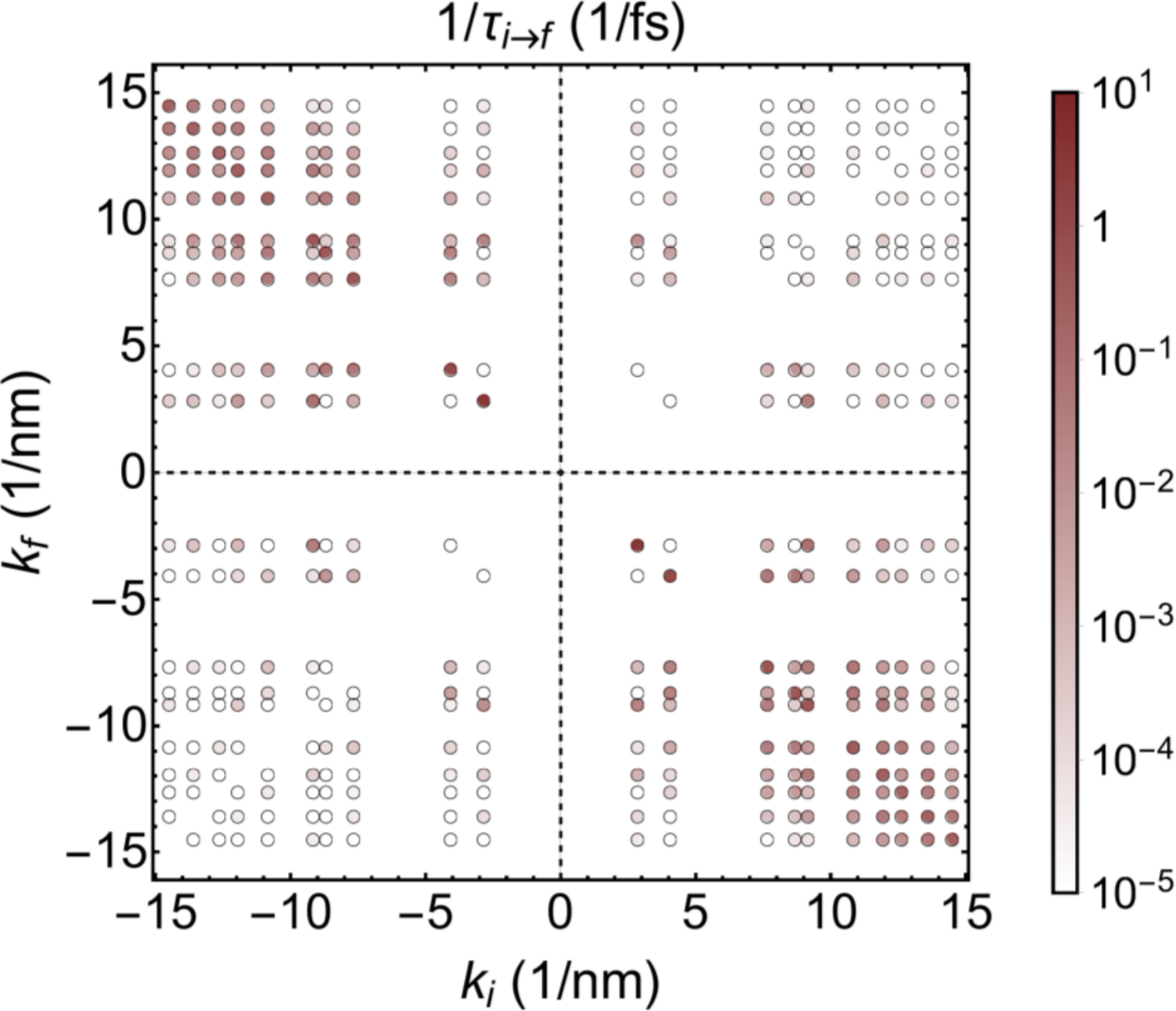}} \hfill
	\subfigure[]{\includegraphics[width=0.49\linewidth]{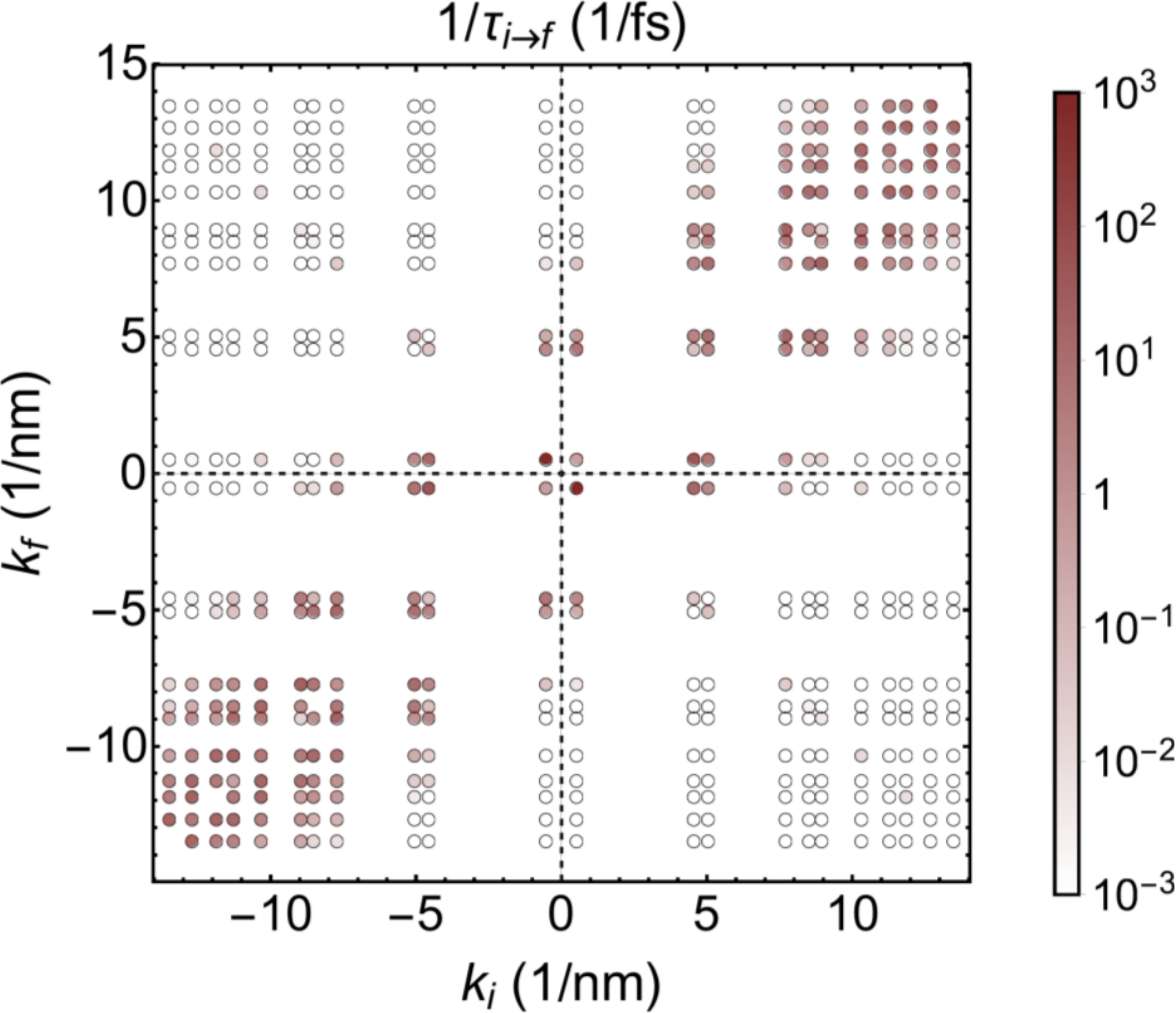}}
	\caption{The scattering rates between specific initial and
	final states are shown as a function of the wave number of
	the initial ($k_i$) and final ($k_f$) states for (a) grain
	boundary scattering (b) surface roughness scattering. A toy
	model system, with a limited number of subbands (or Fermi
	level states), is considered here for clarity, showing the
	important qualitative features.}
	\label{fig:Scatter}
\end{figure}

In Fig.~\ref{fig:Scatter} we show the scattering rates between
pairs of initial ($i$) and final ($f$) states for grain boundary
and surface roughness scattering. The two scattering mechanisms
show very different behavior, the highest grain boundary and
surface roughness scattering rates being concentrated along the
anti-diagonal and diagonal in the $(k_i,k_f)$-plane respectively.
We have included some additional averaging over random
orientation of the grain boundary planes to obtain more realistic
grain boundary scattering rates. This leads to deviations from
perfect anti-diagonal coupling (corresponding to $k$ to $-k$
backscattering) which follows from the standard Mayadas-Shatzkes
expression \cite{moors2015electron}.

\subsection{Resistivity}
\begin{figure}[tbh]
	\centering
	\subfigure[]{\includegraphics[width=0.485\linewidth]{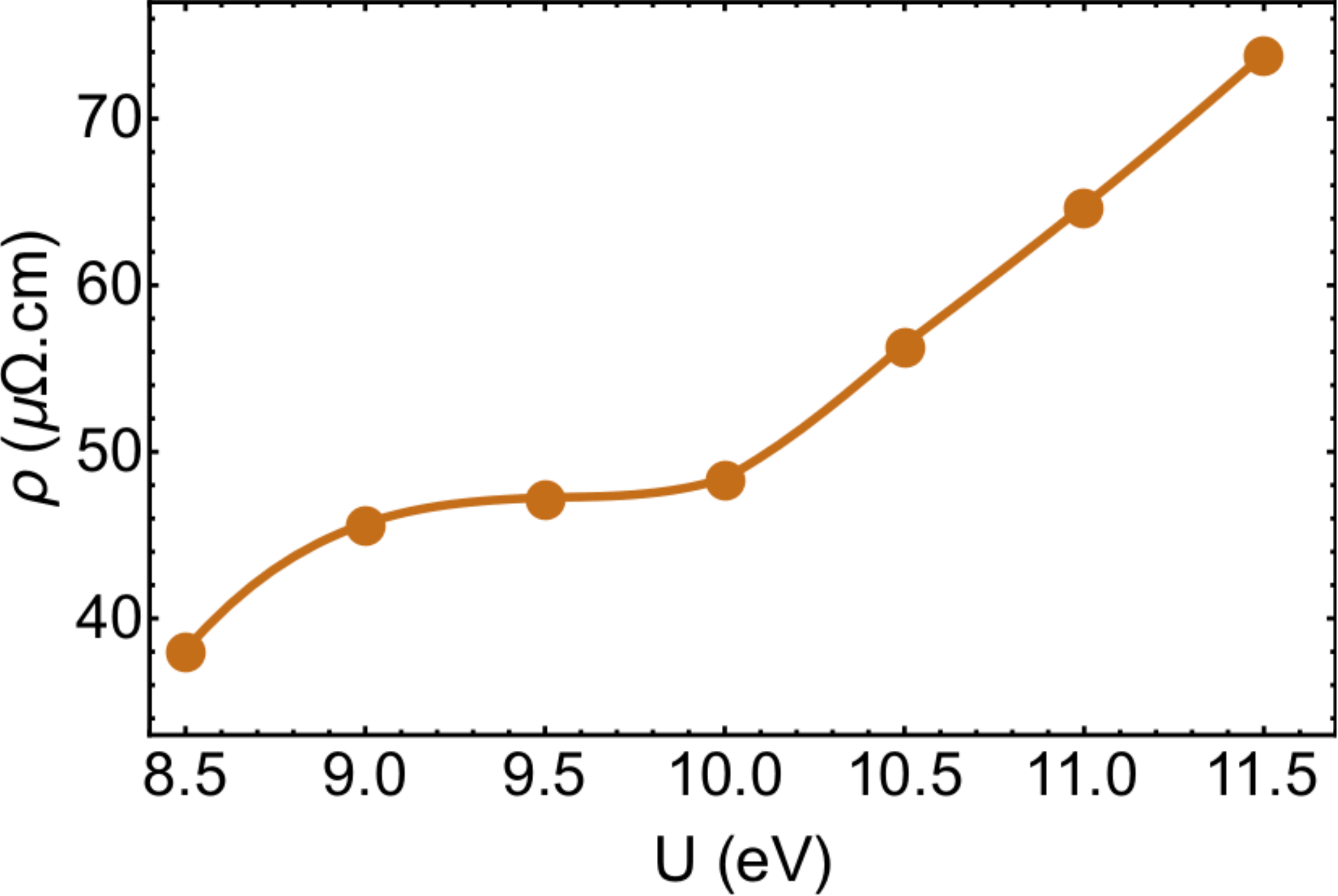}} \hfill
	\subfigure[]{\includegraphics[width=0.495\linewidth]{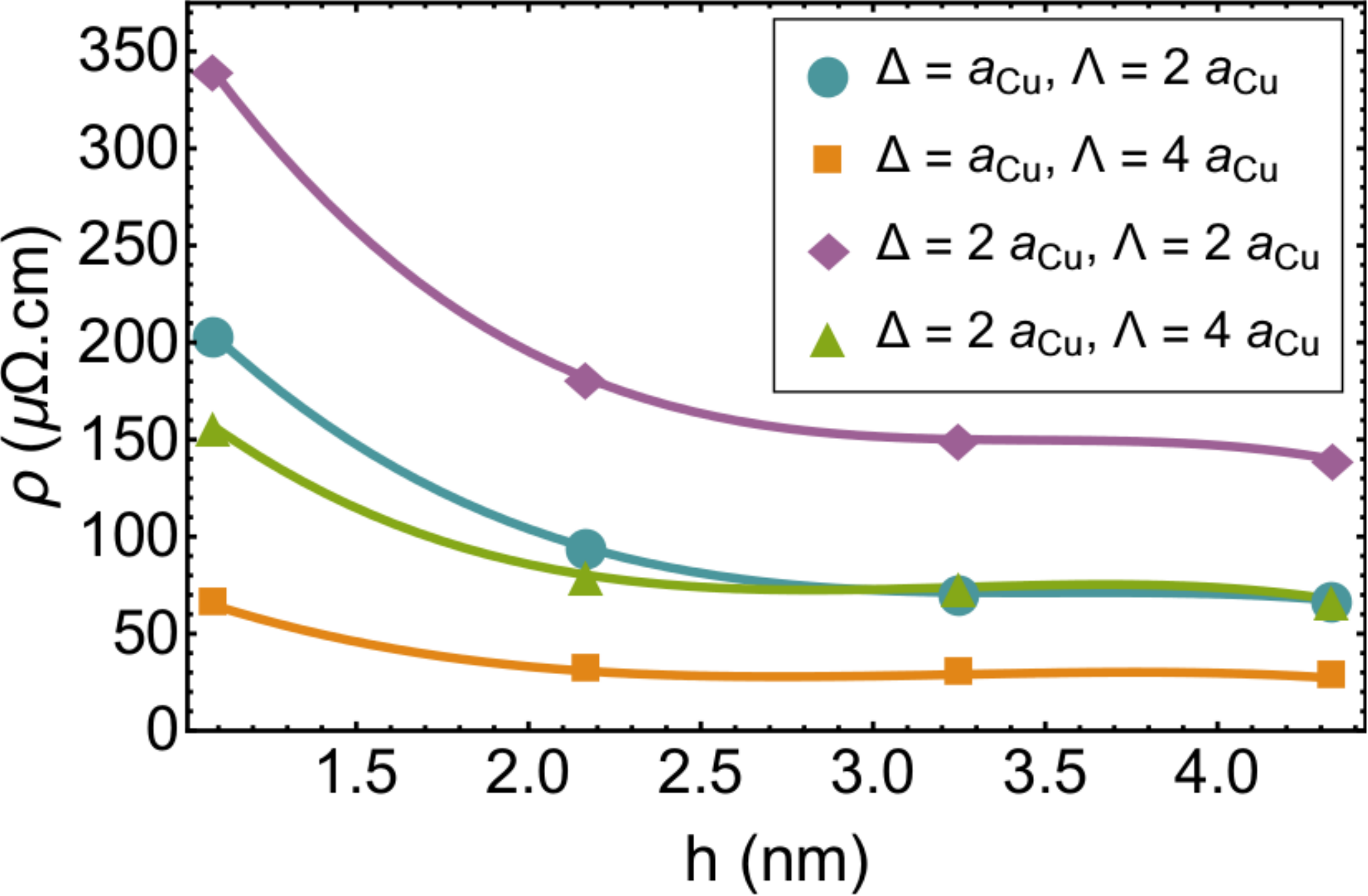}}
	\caption{The resistivity for rough metallic thin films is shown
	as a function of (a) the barrier potential height $U$ for a film
	with thickness $h=12 \, a_\textrm{Cu}$, roughness RMS $\Delta =
	a_\textrm{Cu}$ and correlation length $\Lambda = 2 \,
	a_\textrm{Cu}$ (b) the film thickness $h$ with different surface
	roughness RMS and correlation length values. The vacuum potential
	barrier height is taken to be $U=11.5$~eV.}
	\label{fig:Thin_Film}
\end{figure}

The resistivity for thin films is presented in Fig.~\ref{fig:Thin_Film}
as a function of the potential well barrier height and film thickness.
In general, the resistivity increases with increasing barrier height or
roughness RMS, and with decreasing roughness correlation length or film
thickness. A maximal barrier height is obtained for a vacuum barrier
and can be extracted from the work function $W$.

\begin{figure}[tbh]
	\centering
	\subfigure[]{\includegraphics[width=0.462\linewidth]{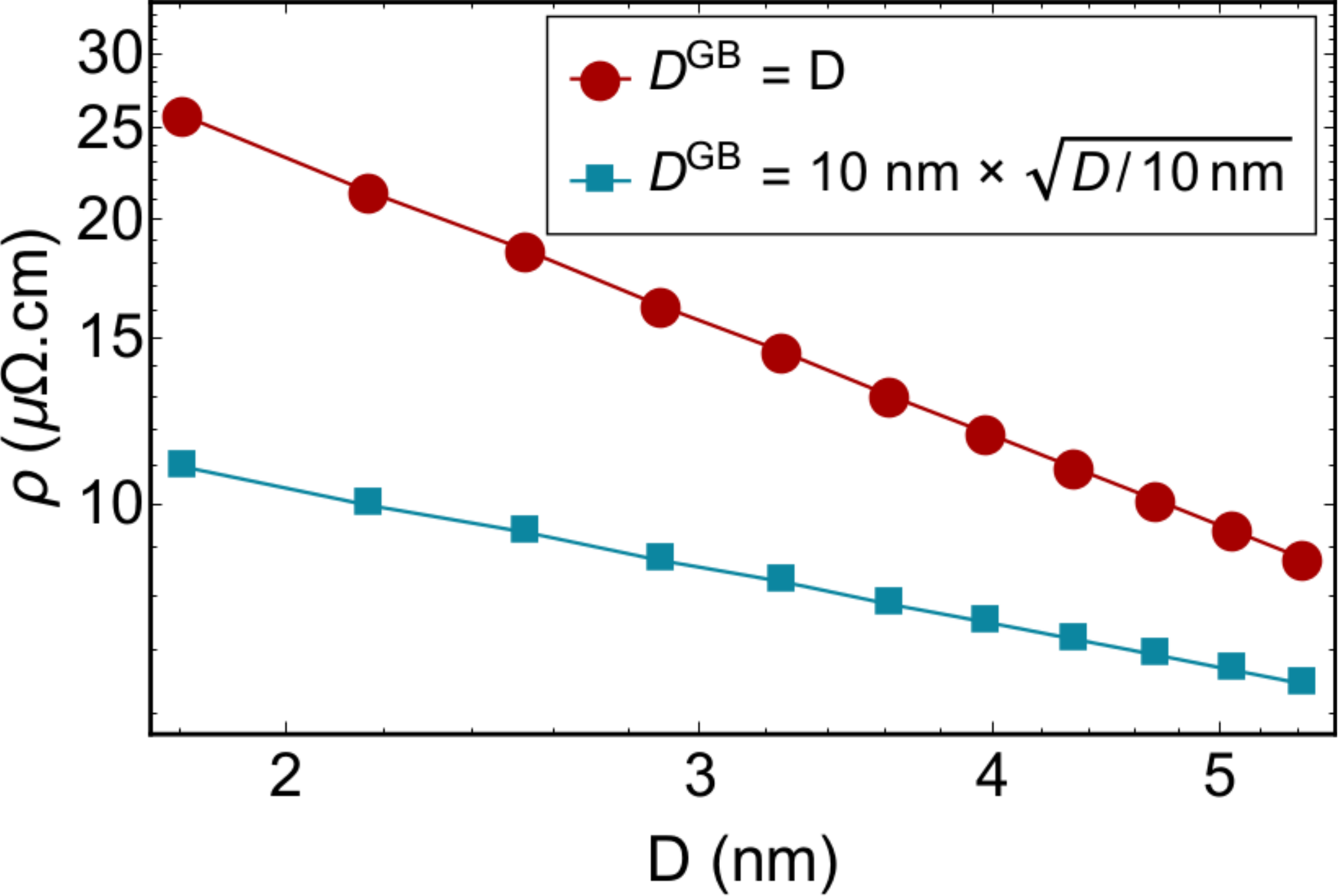}} \hfill
	\subfigure[]{\includegraphics[width=0.518\linewidth]{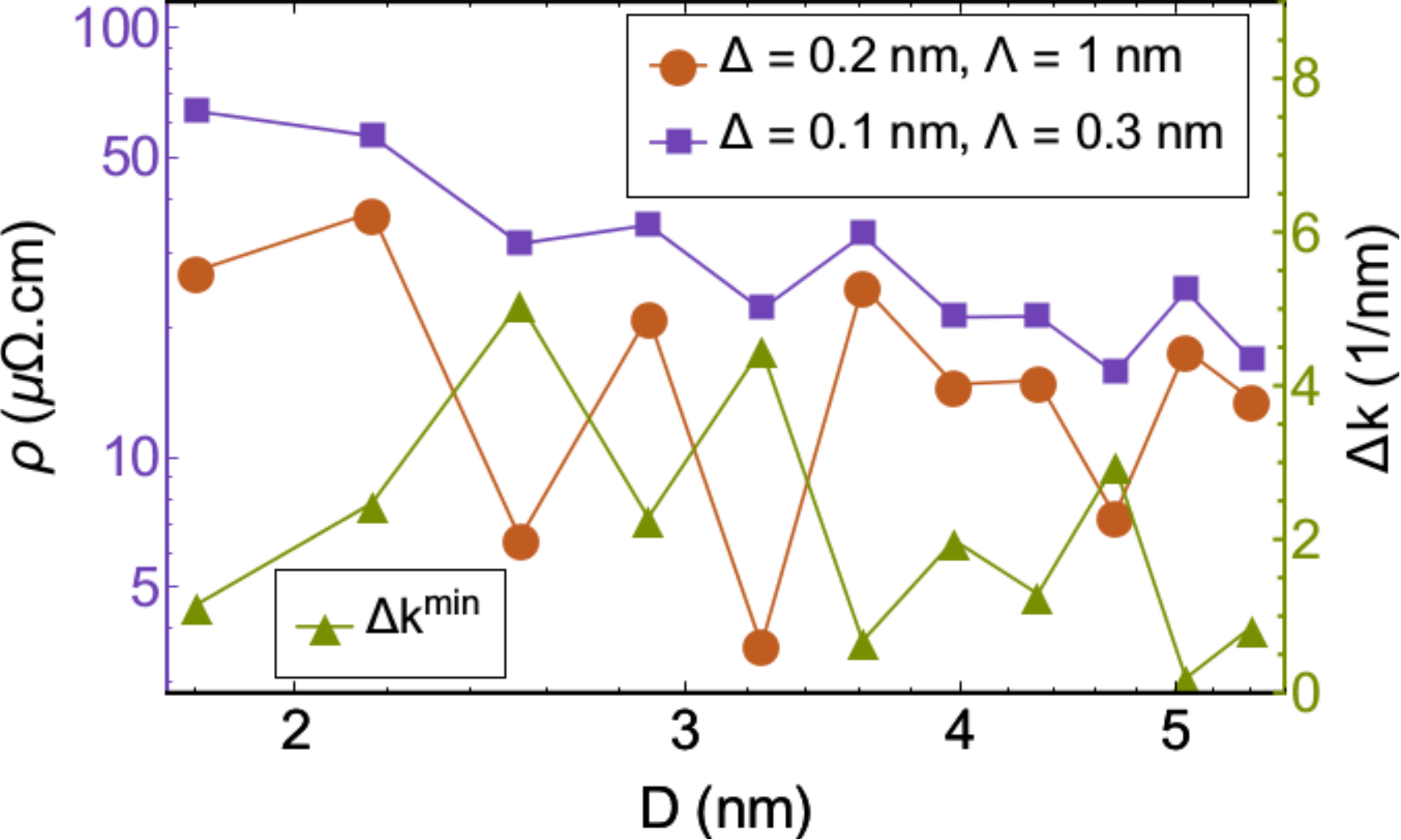}}
	\caption{The resistivity for metallic nanowires with (a) grain
	boundaries (b) boundary surface roughness is shown as a function
	of the nanowire diameter $D$. (a) Two different relations between
	the wire sides and the average inter-grain boundary distance are
	considered: linear and sublinear, with grain boundary strength
	$U^\textrm{GB} = a_\textrm{Cu} \times 1.5$~eV. (b) Two different
	roughness profiles are shown, together with the minimal wave
	number difference $\Delta k^\textrm{min}$ between Fermi level
	states with positive and negative wave numbers for each
	simulated diameter. A finite potential well with barrier
	height $U = E_\textrm{F} + W \approx 11.5$~eV is considered to
	represent vacuum.}
	\label{fig:Nanowire}
\end{figure}

Fig.~\ref{fig:Nanowire} exhibits the resistivity of nanowires as a function
of their side lengths and for various grain boundary or surface roughness
properties. We consider a square cross section for the nanowires and refer
to the side length as the diameter. For grain boundaries we consider a
linear relation between the diameter and the average inter-grain boundary
distance as well as a sublinear relation. The result are very similar to
those of thin films with the resistivity scaling purely determined by the
inter-grain boundary distance and no visible additional effects of
confinement. The standard deviation $\sigma^\textrm{\tiny GB}$ is not
studied, because as long as it is not substantially smaller than
$D^\textrm{\tiny GB}$, resembling an unlikely periodic grain superlattice
structure, its impact on the resistivity is negligible.

Surface roughness is studied for two cases, with different values for both
the standard deviation and correlation length. A general trend of
increasing resistivity for smaller diameters is observed, but there is no
clear scaling exponent and large resistivity drops appear for certain
diameters in case of sufficiently large roughness correlation lengths.
These drops correspond to nanowires with a large minimum of thewave number
difference $\Delta k^\mathrm{min}$ between Fermi level states with positive
wave numbers and their negative counterparts (see Fig.~\ref{fig:Bands}).

\section{Discussion} \label{section:discussion}
We have extracted useful information from the scattering rates of
grain boundary and surface roughness scattering presented in
Fig.~\ref{fig:Scatter} and the simulation results for resistivity
scaling of thin films (Fig.~\ref{fig:Thin_Film}) and nanowires
(Fig.~\ref{fig:Nanowire}). Grain boundaries mostly induce
backscattering which is barely affected by increasing confinement
that accompanies shrinking side lengths. Hence, the resistivity
scaling behavior is similar for thin films and nanowires and
depends on the grain boundary strength and density. The average
grain size and corresponding inter-grain boundary distance are
equally crucial for thin films and nanowires and should be
maximized for an optimal resistivity. It should be noted that
the above results are obtained within the effective mass
approximation. Consequently, when a more realistic band structure
is adopted, together with more realistic grain boundary
potentials, the scattering probability rates may be altered.
But as the grains and their boundaries are typically randomly
distributed and oriented throughout the structure, a significant
suppression of grain boundary backscattering is generally not
expected.

Boundary surface roughness causes very different scattering
behavior as it mostly leads to scattering events with small
scattering angles. While the corresponding scattering rates
can be very large, there is no substantial loss of current as
the transport velocity of the electrons is barely affected.
Loss of current occurs largely through scattering events between
states that have a wave number close to $k=0$. This gives rise
to typical resistivity scaling behavior for thin films, its
resistivity value depending on the barrier height and specific
roughness RMS and correlation length values, where we observe
a non-quadratic relation between resistivity and barrier height
due to the non-linear treatment of the surface roughness
function. For nanowires however, drops in resistivity appear
for certain diameters. These drops coincide with the absence
of Fermi level states close to $k=0$. One can quantify this
absence by looking at the minimal wave number difference
$\Delta k^\textrm{min}$ between Fermi level states with
positive $k$ and those with negative $k$. A critical
difference, $\Delta k^\textrm{crit} = \sqrt{8}/\Lambda$, can
be retrieved from the roughness scattering matrix elements
revealing that backscattering is suppressed exponentially
when $\Delta k^\textrm{min} > \Delta k^\textrm{crit}$,
leading to a resistivity drop \cite{moors2015modeling}. This
drop cannot be explained merely in terms of a phenomenological
specularity parameter and requires a scattering description
with quantized transport wave vectors due to confinement and
boundary surface roughness with a finite correlation length.

\section{Conclusion} \label{section:conclusion}
The resistivity scales up drastically when the diameter or
thickness of nanowires and thin films drops below 100 nm.
In the sub-10 nm regime, quantum mechanical effects of
confinement and scattering come into play, introducing
additional complexity for the resistivity scaling behavior.
The simulation results show a general trend of increasing
resistivity when the nanowire side lengths are reduced,
but the typical resistivity scaling that is observed for
larger wires and thin films is not pursued, mainly due to
confinement changing the surface roughness scattering
properties.

The transport model used to obtain the above mentioned
results is based on the semiclassical multi-subband Boltzmann
transport equation, while allowing for fast and accurate
simulations without fitting parameters. These simulations
in turn provide the means to perform a rigorous analysis of
the impact of band structure and barrier properties as well
as grain and roughness statistics on the resistivity scaling
of metallic thin films or nanowires.

\section*{References}

\bibliography{mybibfile}

\end{document}